\documentclass[amsmath,amssymb,aps,prl,twocolumn]{revtex4-2}

\usepackage{graphicx}

\usepackage{comment}
\usepackage{braket}
\usepackage{ifthen}

\newcommand{\rhovac}{\rho_{ \text{vac} }}
\newcommand{\pvac}{p_{ \text{vac} } }
\newcommand{\Tvac}{T_{\text{vac} }^{\mu \nu}}

\usepackage{subfloat}

\usepackage{hyperref} 
\hypersetup{
 colorlinks=true,
 linkcolor=blue,
 citecolor=blue,
 filecolor=blue, 
 urlcolor=blue,
 pdfpagemode=FullScreen,
 }

\newif\ifincludeOutline 
\includeOutlinetrue 

\begin{document}

\title{ Logarithmically Divergent Vacuum Energy in Effective Field Theory}

\author{Jonathan Sullivan-Wood}
 \affiliation{ Department of Physics and Astronomy, Purdue University, West Lafayette, Indiana 47907}
 \email{sulli391@purdue.edu}
\author{Craig Pryor}
\affiliation{ Department of Physics and Astronomy, University of Iowa, Iowa City, Iowa 52242}
 \email{craig-pryor@uiowa.edu}

\date{\today}

\begin{abstract}
The vacuum energy density due to a single quantum field diverges quarticly with the ultraviolet cutoff $\Lambda$, in wild disagreement with the value implied by cosmological observations.
We show that in effective field theories containing bosons and fermions the requirement of a Lorentz invariant vacuum makes any divergence faster than logarithmic exponentially unlikely.
We show this by generating an ensemble of mass spectra by Monte Carlo, and find that the probability distribution function for the vacuum energy is a gaussian centered on zero with a width that grows only logarithmically with the ultraviolet cutoff.

\end{abstract}
\maketitle

{\bf Introduction}
The vacuum energy or cosmological constant problem lies in the stark mismatch between the enormous vacuum energy density, $\rhovac$, predicted by quantum field theory\cite{Weinberg.rmp.1989,Carroll.araa.1992,Carroll.lrr.2001,Peebles.rmp.2003,Martin.crp.2012} and the tiny value inferred from cosmological observations\cite{riess.aj.1998,Perlmutter.aj.1999}. 
Each state of a non-interacting quantum field with mass $m$ and wave vector $\mathbf k$ contributes a zero-point energy $E({\mathbf k}, m)= \pm\frac{1}{2} \sqrt{ {\mathbf k}^2 + m^2 }$, where the $+$ sign applies to bosons, the $-$ sign applies to fermions, and $\hbar = c = 1$.
The discrepancy results from summing the zero-point energies up to a cutoff scale $\Lambda$ resulting in $\rhovac\propto \Lambda^4$.
It is implicitly assumed that physics above the cutoff does not contribute to the vacuum energy, but even the lowest plausible values of $\Lambda$ result in a predicted value many orders of magnitude larger than what is observed.
If $\Lambda$ is the Planck scale, the calculated value of $\rhovac$ is 122 orders of magnitude too large.

Numerous theories have been proposed to address the cosmological constant problem, each with its own strengths and weaknesses.
These have included introducing new symmetries, such as supersymmetry, which can in principle cancel large contributions to the vacuum energy but does not naturally explain the observed small value after supersymmetry breaking \cite{Weinberg.rmp.1989,Carroll.lrr.2001}. Dynamical dark energy models, such as quintessence, posit that the vacuum energy density evolves over time rather than remaining constant \cite{Peebles.rmp.2003}. Anthropic arguments suggest that the observed value of the cosmological constant is a consequence of selection effects in a multiverse, where only universes with small vacuum energy can support life \cite{Weinberg.prl.1987,Martel.apj.1998,Hogan.rmp.2000, Vilenkin.ijtp.2003}. Despite these efforts, a fully satisfactory and universally accepted solution remains elusive \cite{Martin.crp.2012}.
 
 Pauli was the first to note that bosons and fermions contribute to $\rhovac$ with opposite signs, raising the possibility of cancellations\cite{Pauli.lec.1971}. 
Pauli assumed free fields and $\Lambda \rightarrow \infty$ to obtain a set of conditions on the mass spectrum in order for $\rhovac$ to be finite or zero.
Such cancellations would seem to require an extreme degree of fine tuning, however it has been shown that the finiteness of $\rhovac$ is equivalent to requiring a locally Lorentz invariant vacuum.
That is, the vacuum energy-momentum tensor must be proportional to the Lorentzian metric, \( \Tvac = \langle 0 | \hat{T}^{\mu \nu} |0 \rangle \propto \eta_{\mu \nu} \) \cite{Akhmedov.xxx.2002,Koksma.xxx.2011,Visser.particles.2018a}.
The seemingly unnatural tuning is then a consequence of a natural symmetry, although the finite value of $\rhovac$ remains arbitrary and depends on the details of the mass spectrum.
A number of approaches have been proposed to eliminate Lorentz-violating divergences from a single field by modifying the regulator \cite{Akhmedov.xxx.2002, Ossola.epjc.2003} or reinterpreting the divergences\cite{Koksma.xxx.2011}.
Others have investigated what additional particles could be added to the Standard Model to obtain a small $\rhovac$ \cite{Alberghi.jetpl.2008, Anselmi.epjc.2010, Anselmi.prd.2009,Visser.plb.2019}.
 
 In lieu of a more fundamental theory, physical phenomena are captured by an effective field theory (EFT)\cite{Georgi.arnps.1993,Ebert.ppnp.1994,Brivio.pr.2019,Scherer.book.2011,Hammer.rmp.2020,Meng.pr.2023} whose predictive power only holds up to a finite energy scale \(\Lambda\), not \( \Lambda \rightarrow \infty \). 
Within this framework, \(\rhovac \) is determined entirely by the mass spectrum of the EFT, which reflects the integrated-out effects of higher energy physics that remains inaccessible. 
This opens the door to a more general inquiry: how does \(\rhovac \) behave within the space of EFTs whose spectra are compatible with a Lorentz-invariant vacuum?

One complication is that $\frac{1}{2} \sqrt{ \mathbf{k}^2 + m^2 }$ is the zero-point energy of a non-interacting field. 
While the effects of electromagnetic and weak interactions can be systematically incorporated using perturbation theory \cite{kastening.prd.1996, Kastening.prd.1998}, this is not suitable for strongly interacting quarks and gluons which do not exist as free states at low energies. 
Non-perturbative contributions from QCD remain difficult to compute and introduce significant uncertainties in the estimation of $\rhovac$\cite{Gogohia.prd.2000, Barnafoldi.jpg.2010}.
Alternatively, their contributions to the zero-point energy can be accounted for indirectly via the spectrum of hadronic bound states in which they are dynamically realized.
This replaces the challenge of directly computing the interacting $\rhovac$ with the task of determining bound state masses, a worthwhile trade-off since mass spectra are measurable. 

To explore these ideas we consider $\rhovac$  under a rather broad set of assumptions:\\
(1) The low energy physics is described by an EFT with bound states that interact so weakly they are well approximated as non-interacting.\\
(2) The EFT is valid up to a cutoff scale $\Lambda$.\\
(3) The mass spectrum gives rise to a locally Lorentz invariant vacuum state.\\
Since the imposition of Lorentz invariance turns out to be rather complicated, we investigate this model numerically by generating an ensemble of masses adjusted to satisfy Lorentz invariance.
Each such mass spectrum results in a $\rhovac$, and the ensemble provides a probability distribution function (PDF) for $\rhovac$ over the space of EFTs.

{\bf Zero-Point Energy}
With a finite cutoff scale \(\Lambda\), the vacuum energy is 
\begin{widetext}
\begin{align}
	\rhovac 
 &= \sum_n (-1)^{2 s_n} g_n \int_0^{\sqrt{\Lambda^2-m_n^2}} ~ \frac{d^3k}{(2\pi)^3} \frac{1}{2} \sqrt{ k^2 + m_n^2}\\
 &= \sum_n (-1)^{2 s_n} g_n\left[ {\Lambda \sqrt{\Lambda^2-m_n^2} \left(2 \Lambda^2-m_n^2\right) +m_n^4 \ln
 \left(\frac{-\sqrt{\Lambda^2-m_n^2}+\Lambda}{m_n}\right)}\right] / {32 \pi^2}
 \label{eq:rho_vac_eft} \\
	&= \sum_n (-1)^{2 s_n} \frac{g_n}{16\pi^2}\left[\Lambda^4-m_n^2\Lambda^2+\frac{m_n^4}{8}+\frac{m_n^4}{2} \ln\left( \frac{m_n}{2\Lambda} \right)+\mathcal O(1/\Lambda^2) \right] \label{eq:rho_vac_approx}
 \end{align}
 \end{widetext}
 where $n$ labels the particle species with mass $m_n$, spin $s_n$, and degeneracy $g_n$ due to spin, polarization, or internal degrees of freedom \cite{Visser.particles.2018a}. 
The large-$\Lambda$ expansion in the last line provides Pauli's conditions for $\rhovac$ to be finite as $\Lambda\rightarrow \infty$,
\begin{align}
0= \sum_n (-1)^{2s_n}g_n 
 = \sum_n (-1)^{2s_n}g_n m_n^2 
 = \sum_n (-1)^{2s_n}g_n m_n^4. \label{eq:pauli}
\end{align}
If these conditions are satisfied, then the remaining finite term gives
\begin{align}
\rhovac =\sum_n (-1)^{2s_n}g_n \frac{1}{64\pi^2} m_n^4 \ln \left( m_n^2 / \mu^2 \right) \label{eq:rho_vac_inf}
\end{align}
where $\mu$ is an arbitrary scale. 
Since Eq.~\ref{eq:pauli} is assumed to be satisfied, $\rhovac$ will be independent of $\mu$.
Pauli's original conditions were Eq.~\ref{eq:pauli} plus setting Eq.~\ref{eq:rho_vac_inf} to zero since that was consistent with observations at the time.

Local Lorentz invariance is dictated by the structure of the zero-point energy-momentum tensor\cite{Visser.particles.2018a}
\begin{align}
\Tvac
 &= \text{diag} ( \rhovac , \pvac, \pvac, \pvac ) 
\end{align} 
where the zero-point pressure $\pvac$ is given by
\begin{widetext}
 \begin{align}
	\pvac &= \sum_n (-1)^{2 s_n} g_n \int_0^{\sqrt{\Lambda^2-m^2}} \frac{d^3k}{(2\pi)^3} \frac{1}{6} \frac{k^2}{ \sqrt{ k^2 + m_n^2} } \label{eq:p_vac_eft} \\
 &=\sum_n (-1)^{2 s_n} g_n \left[{\Lambda \left(2 \Lambda^2-5 m_n^2\right) \sqrt{\Lambda^2-m_n^2}+3 m_n^4 \ln \left( \frac{\Lambda+\sqrt{\Lambda^2-m_n^2}}{m_n}\right)} \right] /{96 \pi ^2} \label{eq:p_vac_eft}\\
	&=\sum_n (-1)^{2 s_n} g_n \left[ \frac{\Lambda^4}{48\pi^2} -\frac{\Lambda^2 m_n^2}{16\pi^2} +m_n^4\frac{3+4\ln\left( \frac{2\Lambda}{m} \right) }{128\pi^2}+ {\mathcal O}(1/\Lambda^2) \right]. \label{eq:p_vac_approx}
\end{align}
\end{widetext}
In order to be Lorentz invariant, $\Tvac$ must be proportional to the Minkowski metric, or $\pvac = - \rhovac$.
The sum rules in Eq.~\ref{eq:rho_vac_inf}, which cancel the quartic and quadratic divergences in $\rhovac$ also eliminate the divergences in $\pvac$, leaving $\pvac = -\rhovac$.
Thus, finiteness of $\Tvac$ is equivalent to a Lorentz invariant vacuum state\cite{Visser.particles.2018a}.

{\bf Effective Field Theory}
An EFT is defined by a cutoff $\Lambda$ and a set of coupling constants that are determined by matching conditions, typically involving observables such as scattering amplitudes or decay rates\cite{Georgi.arnps.1993,Burgess.arnps.2007,Weinberg.book.2016,Hammer.rmp.2020,Meng.pr.2023}. 
Such conditions aren't applicable to the vacuum state, but we can use Lorentz invariance by adding Eq.s \ref{eq:rho_vac_eft} and \ref{eq:p_vac_eft} to obtain
\begin{equation}
0 = \rhovac+\pvac = \frac{\Lambda}{12\pi^2} \sum_n (-1)^{2 s_n} g_n(\Lambda^2-m_n^2)^{\frac{3}{2}} \label{eq:Lorentz_EFT} 
\end{equation}
which must be obeyed by the mass spectrum of an EFT.
Therefore the vacuum energy problem should be framed as asking what tuning is required to obtain the observed $\rhovac$ in an EFT that obeys Eq. \ref{eq:Lorentz_EFT}.

We investigate this problem numerically by generating random mass spectra and then adjusting  to impose Eq.~\ref{eq:Lorentz_EFT}.
The histogram of the computed $\rhovac$ values yields the PDF of $\rhovac$ within the space of physically allowed spectra.
This distribution allows us to assess how typical or how finely tuned a value of $\rhovac$ is, offering a  test of anthropic reasoning: if the observed value lies near the peak of the distribution, it may be considered statistically natural under mild selection criteria \cite{Weinberg.prl.1987, Martel.apj.1998}; if it lies in a tail, stronger anthropic constraints or selection mechanisms may be required \cite{Bousso.jhep.2000, Garriga.prd.2000, Douglas.universe.2019}.
To obtain an ensemble of mass spectra, we need an initial PDF with which to generate the masses.
We will be guided by the qualitative features of the measured spectrum, but our results will be seen to be independent of the underlying mass PDF.
The two central considerations guiding our choice are the role of interactions and particle lifetimes.

As mentioned earlier, we address the problem of interactions by evaluating $\rhovac$ using the quasi-free bound states for which Eq.s \ref{eq:rho_vac_eft} and \ref{eq:p_vac_eft} are justified.
There is a long history of statistically modeling the energy levels of strongly interacting systems going back to Wigner's random matrix theory for nuclei\cite{wigner.siamr.1967,Edelman.an.2005}. 
In the 1960s, Hagedorn noted that the number of hadronic states increases rapidly with mass and is fit by a growing exponential of the form ${\mathcal P}(m) \propto \exp( m / T_H)$ \cite{Hagedorn.ncs.1965}, where $T_H\approx 200 ~\rm MeV$ is fit to the experimental data \cite{Hagedorn.ncs.1965}.
As more data have become available, the trend has continued \cite{Bugaev.el.2009,Broniowski.plb.2000,Broniowski.prd.2004, Cohen.jpg.2012} and the Hagedorn spectrum has also been seen in lattice QCD calculations \cite{Lo.prc.2015,Alba.prd.2017}.
The number of quasi-free fields, such as leptons, is relatively small compared to the exponentially large number of hadrons. 
As a result, adding a few extra particles would be difficult to distinguish from statistical fluctuations within the hadronic spectrum and a single exponential distribution may be used.
In any case, we find the final results do not depend on this exponential distribution.
For the known particles, lifetimes decrease with increasing mass so the masses contributing to $\rhovac$ will be limited to those for which the decay width obeys $\Gamma_n \ll m_n$.
The effects of lifetimes are modelled by imposing a mass cutoff $M$.

\begin{figure}[ht!] 
	 \centering
	 \includegraphics[width=1.0\linewidth]{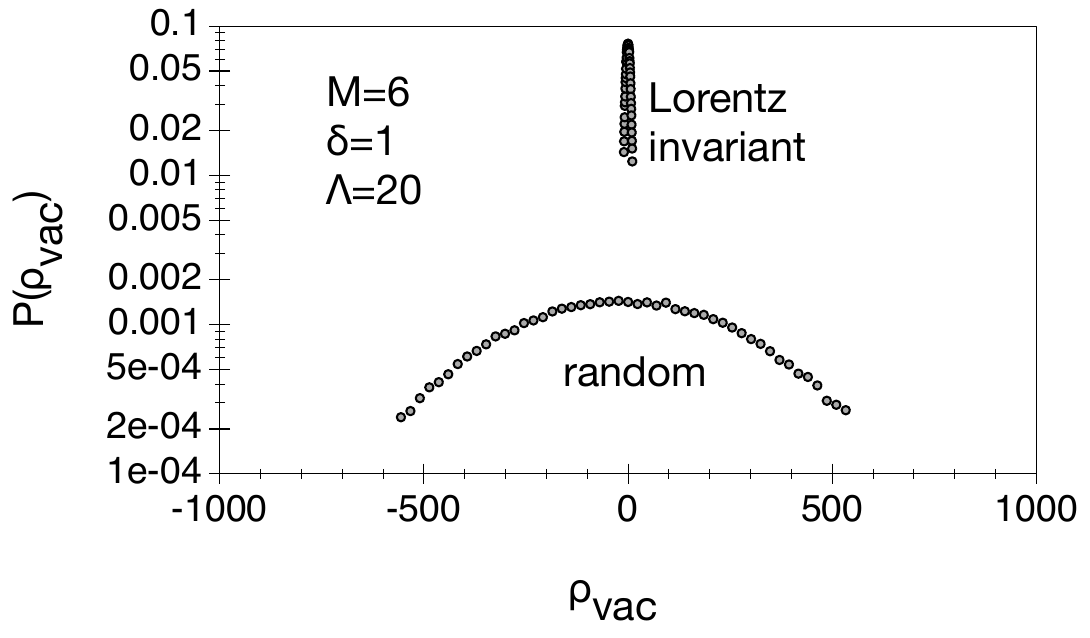} 
	\caption{Probability distributions functions for $\rhovac$ before and after imposing Lorentz invariance on the mass spectrum, with $T_H =1$, cutoff $ \Lambda  =20$, mass cutoff $M=6$, and  average minimum mass $\delta= 1$.
	The broad curve is from $10^5$ unconstrained random mass spectra generated and the narrower curve is the distribution resulting from the imposition of $\rhovac = -\pvac$. Both are gaussians centered on $\rhovac=0$. } \label{fig:Lorentz_narrowing}
	\end{figure}
	\begin{figure}[ht!]
	\centering
	\includegraphics[width=1.0\linewidth]{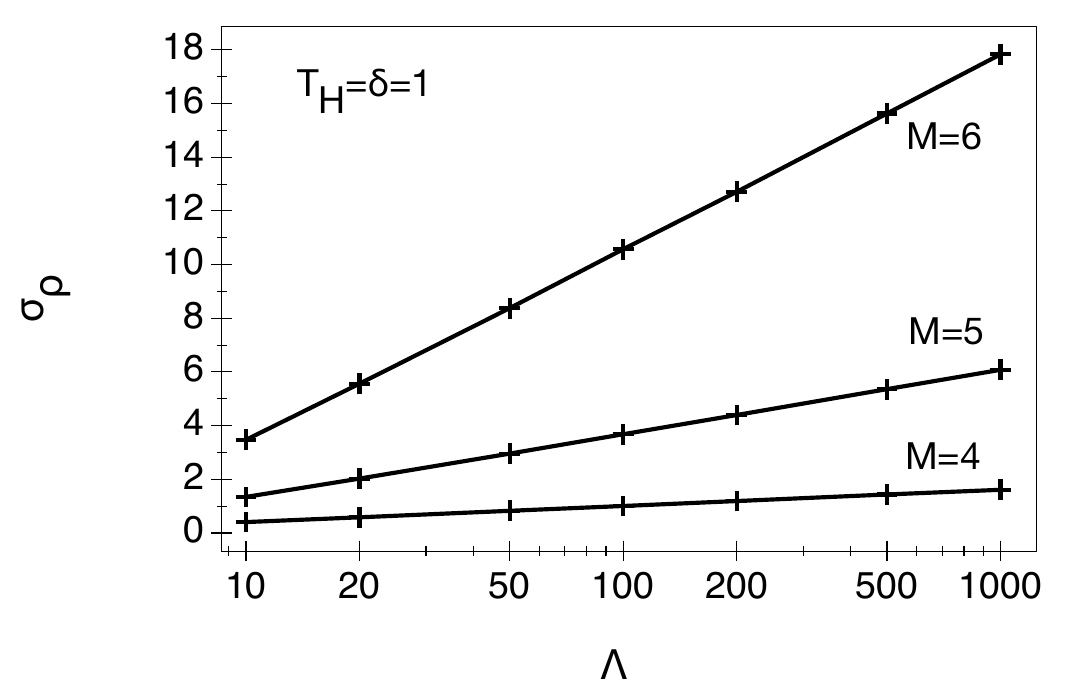}
	\caption{ Standard deviation of the distribution of $\rhovac$ as a function of the cutoff $\Lambda$ for different mass cutoffs $M$.
	Each point was computed using an ensemble of $10^5$ mass spectra,  with $T_H=1$ and the indicated $M$. 
	The number of masses in each spectrum (varying from 22 to 276) was chosen to the make the ensemble average of the minimum mass $\delta =1$. 
	Points are connected with straight lines to show the logarithmic dependence on cutoff.
	}
	\label{fig:scaling}
\end{figure}

{\bf Results}
Masses were generated with probability ${\mathcal P}(m) \propto \exp ( m / T_H) $ up to the mass cutoff $M$, with equal numbers of bosons and fermions, and degeneracy factors set to $g_n=1$.
The number of masses was adjusted so as to make the ensemble average of the lowest mass equal to a specified value $\delta$, which we set to 1.
Lorentz invariance was imposed by randomly selecting a mass and attempting to adjust it to satisfy Eq.~\ref{eq:Lorentz_EFT}. 
This was attempted up to 10 times, and if unsuccessful the spectrum was discarded and a new one was generated.
We generated ensembles of mass spectra with $T_H=1$ and $\delta=1$ while varying the maximum mass $M$ and the cutoff $\Lambda$.

Since Lorentz invariance was enforced by tuning only a single mass, the resulting spectra remained statistically close to the original distribution. 
Although the subtle mass correlations required by Lorentz invariance leave the overall distribution largely unchanged, they substantially affect $\rhovac$.
As seen in Fig.~\ref{fig:Lorentz_narrowing} the PDF for $\rhovac$ from unconstrained mass spectra is a gaussian centered on zero.
In the limit of an infinite number of masses this would be guaranteed by the central limit theorem.
The imposition of Eq.~\ref{eq:Lorentz_EFT} introduces correlations which significantly narrows the distribution of $\rhovac$, but it remains gaussian.

The standard deviation $\sigma_\rho$ quantifies the variation in $\rhovac$ and serves as an indicator of its expected dependence on $\Lambda$ across the ensemble of EFTs.
Fig.~\ref{fig:scaling} shows $\sigma_\rho$ calculated from different values of $\Lambda$.
Remarkably, the results are well fit by a logarithmic dependence on cutoff.
This illustrates that Lorentz invariance, rather than requiring finely tuned cancellations, naturally suppresses vacuum energy contributions.
The resulting sensitivity to high-energy physics is substantially weaker than what standard estimates would suggest.
Comparing $\sigma_\rho(\Lambda)$ for different values of $M$ we notice an exponential dependence.
This simply reflects the exponentially growing number of masses, and $\rhovac$ scales linearly with the number of masses in the EFT.
We have also done calculations using uniform mass distributions and find the same $\ln \Lambda$ dependence, showing our result is insensitive to the underlying mass distribution.

{\bf Conclusions}
Effective field theories with random mass spectra have a distribution of vacuum energy density that is a gaussian with a peak at $\rhovac=0$.
If the masses are constrained to give a Lorentz invariant vacuum then the distribution of the vacuum energy has a width $\sigma_\rho \propto \ln \Lambda$.
Therefore, a Lorentz invariant EFT is most likely to diverge logarithmically  with the cutoff.
Large values corresponding to $\rhovac \sim \Lambda^4$ are of course possible from the tail of the distribution.
However, out of the possible mass spectra, those with such strong divergence are extremely unlikely due to the normal distribution of $\rhovac$.
Thus, we find that fine tuning is required to obtain {\it large} values of $\rhovac$.
When viewed through the lens of an ensemble of possible Lorentz invariant theories, the discrepancy between theory and observation is reduced from a factor of $\sim 10^{122}$ to a less disturbing $\mathcal{O}(10^2)$.
The sign of $\rhovac$ remains indeterminate, with positive and negative values equally probable -an anthropic coin flip that, while arbitrary, seems benign.
Finally, it is noteworthy that Lorentz invariance is the minimum possible requirement that can be imposed on the vacuum.
If additional constraints reduce the range of $\rhovac$ to something growing more slowly than $\ln \Lambda$, $\rhovac$ might be calculable.

This research was supported in part through computational resources provided by The University of Iowa.

\bibliographystyle{apsrev4-2}
\bibliography{highEnergy.bib} 

\end{document}